\documentclass{article}
\usepackage{spconf,amsmath,graphicx}
\usepackage{amsmath,amsfonts,amssymb,pifont}
\usepackage{adjustbox}
\usepackage{comment}
\usepackage[para]{footmisc}
\usepackage[utf8]{inputenc}

\title{Synergy between human and machine approaches to sound/scene recognition and processing: an overview of ICASSP special session}

\name{Laurie M. Heller$^{1^*}$, Benjamin Elizalde$^{2^*}$, Bhiksha Raj$^{3,4^*}$, Soham Deshmukh$^2$  \vspace{-8.5pt}}
\address{$^*$Special session co-organizers \\$^1$ Department of Psychology, Carnegie Mellon University \\$^2$Microsoft \\ $^3$Languaged  Technologies Institute, Carnegie Mellon University \\$^4$Mohammed bin Zayed University of AI \\ 
laurieheller@cmu.edu, bhikshar@andrew.cmu.edu, \{benjaminm,sdeshmukh\}@microsoft.com\vspace{-8pt}}

\begin{document}

\maketitle

\begin{abstract}
Machine Listening, as usually formalized, attempts to perform a task that is, from our perspective, fundamentally human-performable, and performed by humans. Current automated models of Machine Listening vary from purely data-driven approaches to approaches imitating human systems. In recent years, the most promising approaches have been hybrid in that they have used data-driven approaches informed by models of the perceptual, cognitive, and semantic processes of the human system. Not only does the guidance provided by models of human perception and domain knowledge enable better, and more generalizable Machine Listening, in the converse, the lessons learned from these models may be used to verify or improve our models of human perception themselves. This paper summarizes advances in the development of such hybrid approaches, ranging from Machine Listening models that are informed by models of peripheral (human) auditory processes, to those that employ or derive semantic information encoded in relations between sounds. The research described herein was presented in a special session on ``Synergy between human and machine approaches to sound/scene recognition and processing" (ICASSP 2023).

\end{abstract}

\section{Introduction}
Human auditory knowledge has been used to improve many areas of machine listening and signal processing (Virtanen et al~\cite{virtanen2018computational}, Lyon~\cite{lyon2017human}, Blauert~\cite{blauert1997spatial}, Bowen~\cite{Bowen}), such as: sound event classification, sound synthesis, speech recognition, and binaural and spatial sound processing. These synergies have impacts in human-oriented applications such as alerting people to important sounds, perceptual and cognitive assessment, hearing aids, and spatializing sounds in virtual reality. While the objective of machine listening system is to eventually achieve superhuman performance, the benchmark remains human performance.  Human performance is the emergent outcome of many biological and cognitive processes, from the early-stage processing of the auditory signal, to the late-stage semantic processes that drive our interpretations. 
Automated models of machine listening system may take one of three routes:
\begin{itemize}
    \item A purely data-driven approach that has no explicit reference to the intermediate perceptual and cognitive processes employed by the human;
    \item An imitation approach that attempts to model and mimic the human process in detail; or
    \item A hybrid approach that uses models of the perceptual, cognitive, and semantic processes of the human system to inform a data-driven approach.
\end{itemize}

Of these, the hybrid approach remains the most promising, particularly when combined with the power of recent deep-learning models (cf. Turian et al~\cite{pmlr-v176-turian22a}, Elizalde et al~\cite{Elizalde2021}., Pranay et al~\cite{manocha2020differentiable}, Anderson et al ~\cite{Anderson}, Tashev et al~\cite{lee2015high}, Ananthabhotla~\cite{ananthabhotla2022cognitive}, Wang et al~\cite{wang2017does}, Zeghidour et al~\cite{zeghidour2021leaf}).
We define hybrid approaches as approaches that learn from data-driven methods and use knowledge about human perception and cognition to constrain the optimization problem and make end-to-end learning easier. These methods are usually faster to converge and provide better metrics on the tasks they are solving. This special session on ``Synergy between human and machine approaches to sound/scene recognition and processing" brought together researchers who have worked on a variety of aspects of this hybrid approach, ranging from Machine Listening models that are informed by, or inform models of peripheral (human) auditory processes, to those that employ or derive semantic information encoded in relations between, and response to sounds.  By bringing this group of researchers together, we identified synergies between the data-driven and perception/cognition-based approaches that contribute significantly to advances in both Machine Listening and our knowledge of human auditory perception and cognition.


\section{Hybrid data-driven approaches for human cognition}
The hybrid data-driven approaches explored in the special session can be grouped into two types: (1) models that enhance our capabilities through the use of semantics and perceptual analysis, (2) studies that enhance our understanding of the potential and pitfalls of using data-driven models to assess human listening. 

\subsection{Usage of semantics and perceptual analysis for sound understanding, discrimination, and synthesis}
Cognitive neuroscience research has shown that humans exploit higher-level semantic information about sound sources to understand sound events and infer their context. The approaches learn the semantic information from modality complementary to audio, for example, textual descriptions \cite{elizalde2022clap, deshmukh2022audio} or from available ontology \cite{jimenez2018sound}. In this section, we summarize the findings of research that probes the semantics of audio understanding. 

Esposito et al., in their work entitled ``Semantically-informed deep neural networks for sound recognition" propose SemDNN, a neural network architecture that learns semantic relations from text embeddings (word2vec) while learning sound recognition. They show that SemDNN embeddings approximate human dissimilarity ratings of natural sounds better than those of a traditional (one-hot encoding) sound categorization network (CatDNN). First, they use two evaluation metrics: ranking score and average maximum cosine similarity score (AMCSS). This evaluation was performed for 1 internal dataset and 4 public sound event classification datasets. Second, they compare all the network architectures with human behavioral data using representational similarity analysis (RSA). Overall, they conclude that training with continuous semantic embeddings provides more accurate semantic labelling of sounds and they suggest extending this approach to use different aspects of sound cognition.

Ontologies define concepts and the relation between concepts in a structured form which has semantic meaning. In ``An approach to ontological learning from weak labels," Shah et al. explore using ontological information for learning sound event classifiers. The authors use a graph convolutional neural network (GCN) with an ontological layer for learning the hierarchical structure between sound events. The authors conclude that although GCN as part of the ontology layer captures the ontology knowledge, the model does not perform better by incorporating this ontology information in the weak and multi-labeled sound event classification task.

Thoidis et al., in ``Perceptual analysis of speaker embeddings for voice discrimination between machine and human listening,"  investigate the relationship between machine listening models and human listeners in a speaker verification task. A Convolutional Neural Network (CNN) is trained to conduct one-shot speaker verification. The CNN is trained using a joint loss function which incorporates the cosine distance between the latent features and their corresponding class centers in the penalty of the loss function. The proposed loss function improves one-shot speaker verification performance and makes the network more robust to noise over state-of-the-art approaches. The authors also conduct tests which conclude that a substantial overlap exists between machine and human listening in a voice discrimination task. 

The aim of sound-matching algorithms is to find a set of sound synthesis parameters that minimize the \textit{perceptual} distance between a synthesized sound and its target audio. The current literature uses multiple loss functions ranging from mean square error (known as P-loss) to mean square error in the spectro-temporal domain (known as spectral loss). Han et al., in ``Perceptual neural physical sound matching," devise a Perceptual-Neural-Physical loss (PNP), which is a perceptually-motivated metric that is an approximation of spectral loss which maintains the same training time as spectral loss. Their perceptual similarity metric is an idealized model of spectro-temporal responses in the primary auditory cortex and reflects human judgements. Their PNP loss guides the synthesis of drum sounds using wavelets.


\subsection{Potential and pitfalls of using data-driven models to assess human listening}

Khalil et al., in ``Using machine learning to understand the relationships between audiometric data, speech perception, temporal processing, and cognition," develop a data-driven analysis of perceptual and physiological factors affecting human speech comprehension. The approach uses an ensemble of machine learning models to find the best-performing models that predict the outcomes of three different speech perception tests. The models take 147 features derived from audiometric measurements as inputs. The features also contain new composite variables that represent properties of the entire hearing range. The researchers reported the explanatory power of the different features on the listeners' performance in the three tasks. The prediction models suggest that the mid-frequency range from 1 to 4 kHz is crucial for speech perception since the corresponding features explain most of the variance in the data. In contrast, cognition-related features contribute little to the predictions. Hence the machine learning results serve to remind researchers to sufficiently account for mid-frequency hearing loss when investigating extended high-frequency threshold and cognitive effects on speech perception.

In ``Classifying non-individual head-related transfer functions with a computational auditory model: calibration and metrics,"  Daugintis et al. use a multi-feature Bayesian spherical auditory sound localization model to assess the goodness of non-individual head-related transfer functions (HRTFs) for a human listener. Their template comparison-based model returns a directional probability distribution that is combined with a prior belief. The model is calibrated to individuals, based on their sound localization performance. This paper provides a theoretical framework for a model-based metric that accounts both for acoustic and psychoacoustic similarities in HRTFs. Once perceptually validated, this method could be used as a metric in combination with other methods to enable consistent selection of a well-matched high-quality HRTF. The ultimate goal is to improve the experience of binaural spatial audio technologies.

\section{Discussion}
In this paper, we presented two classes of hybrid approaches. The first approach consisted of four models that enhance our understanding through the use of semantics and perceptual analysis. Two different studies trained deep neural networks with semantic information in the hopes of predicting human data: while the SemDNN showed benefits of using text embeddings for predicting dissimilarity ratings, the ontological learning GCN did not benefit from incorporating sound ontologies for weak multi-labeled sound events. Furthermore, two studies used perceptual analysis to help compare sounds: while speaker embeddings in a CNN improved one-shot speaker verification and agreed with human judgements of voice similarity, a sound-matching algorithm improved a synthesizer of percussive sounds by comparing the target and synthesized sound in terms of a perceptual similarity metric. These four studies demonstrated the potential benefits of using domain knowledge generated by humans to improve machine listening.  

The second hybrid approach consisted of two studies that enhance our understanding of the potential and pitfalls of using data-driven models to assess human listening. First, machine learning models were moderately predictive of the outcomes of speech perception tests and clearly identified the most important variables among those analyzed. Second, a computational model of human sound localization was applied to HRTF evaluation to provide a perceptual foundation for automated HRTF personalization techniques.  


\section{Conclusion}
In conclusion, we demonstrated benefits of a hybrid data-driven approach to machine listening that is informed by models of the perceptual, cognitive, and semantic processes of the human system. Challenges encountered by some of these studies, for example in automatic assessments of human listening, may lead to future improvements. Taken as a whole, these studies demonstrated the potential benefits of using auditory models and domain knowledge generated by humans to improve Machine Listening. 

\bibliographystyle{IEEEbib}

\bibliography{refs}
\end{document}